\documentclass[a4paper,fleqn,usenatbib]{mnras}

\usepackage{color}
\usepackage{graphicx}
\usepackage{amsmath}
\usepackage{amssymb}
\title[Detecting the Elusive Blazar Counter-Jets]{Detecting the Elusive Blazar Counter-Jets}
\author[Liodakis et al.]
{I.Liodakis$^{1,2}$\thanks{liodakis@physics.uoc.gr}, V. Pavlidou$^{1,2}$, and E. Angelakis$^{3}$\\
$^{1}$Department of Physics and ITCP\thanks{Institute for Theoretical
  and Computational Physics, formerly Institute for Plasma Physics}, University of Crete, 71003, Heraklion, Greece\\
$^{2}$Foundation for Research and Technology - Hellas, IESL, Voutes, 7110 Heraklion, Greece\\
$^{3}$Max-Planck-Institut f\"ur Radioastronomie, Auf dem Ḧ\"ugel 69, 53121 Bonn, Germany\\
}
\begin{document}
%\date{Accepted 1988 December 15. Received 1988 December 14; in original form 1988 October 11}
%\pagerange{\pageref{firstpage}--\pageref{lastpage}} \pubyear{2002}
\maketitle
\label{firstpage}
\begin{abstract}
Detection of blazar pc scale counter-jets is difficult, but it can provide invaluable insight into the relativistic effects, radiative processes and the complex mechanisms of jet production, collimation and accelation in blazars. We build on recent populations models (optimized using the MOJAVE apparent velocity and redshift distributions) in order to derive the distribution of jet-to-counter-jet ratios and the flux densities of the counter-jet at different frequencies, in an effort to set minimum sensitivity limits required for existing and future telescope arrays in order to detect these elusive counter-jets. We find that: for the BL Lacs $5\%$ of their counter-jets have a flux-density higher than 100mJy, $15\%$ are higher than 10 mJy, and $32\%$ have higher flux-density than 1 mJy, whereas for the FSRQs $8\%$ have a flux-density higher than 10mJy, $17\%$ are higher than 1 mJy, and $32\%$ are higher than 0.1 mJy (at 15 GHz). Future telescopes like the SKA and newly operating like e-MERLIN and JVLA may detect up to $99\%$ of the BL Lac and $77\%$ of the FSRQ counter-jets. Sources with both low apparent velocity and a low Doppler factor make prime candidates for counter-jet detection. Combining our findings with literature values we have identified five such counter-jet detection candidates. Finally, we discuss possible effects beyond relativistic deboosting that may complicate the detection of counter-jets and that need to be accounted for in the interpretation of detections.
\end{abstract}

\begin{keywords}
galaxies: active -- galaxies: jets -- galaxies: blazars -- galaxies: counter-jet
\end{keywords}

\section{Introduction}\label{introd}

Blazars, i.e Flat Spectrum Radio Quasars (FSRQs) and BL Lac objects, are active galactic nuclei (AGN) with jets oriented close to our line of sight \citep{Readhead1978,Blandford1979,Scheuer1979,Readhead1980}. Their broadband spectrum, from radio to $\gamma$-rays, is characterized by non-thermal radiation and extreme variability. The emission in the low energy regime is dominated by synchrotron radiation of relativistic electrons accelerated in the magnetic field of the jet and thus polarized, whereas the high energy regime is believed to be dominated by the inverse Compton scattering of an ambient low energy photon field. Before the relativistic nature of blazar jets was established, the one-sided morphology was thought by many to  be intrinsic to the source. VLBI observations in the 1970s were critical in forming our current understanding of relativistic extragalactic jets, including the continuity between pc-scale jets and kpc-scale jets and lobes \citep{Readhead1978-II}, their relativistic nature [proposed by   \cite{Rees1966,Rees1967} to reconcile the synchrotron nature of radio emission from compact sources with their rapid variability, and established through the unambiguous detection of superluminal motions \citep{Readhead1978-II}], and the unification between sources viewed at large and small angles with respect to the jet \citep{Readhead1980-II}. The discovery, through spectropolarimetric observations, of the torus surrounding the central regions of radio galaxies \citep{Antonucci1985} completed the modern unification picture of relativistic jets associated with supermassive black holes.

Although theoretical models predict the existence of two symmetrical jets \citep{Rees1966,Blandford1974,Scheuer1974,Readhead1980-II} extending outwards from the central engine, in most cases pc-scale observations of radio jets reveal only one-sided structures \citep{Lister2009}.

\subsection{Doppler favoritism in beamed sources}

Once the relativistic nature of blazar jets was established, it was generally accepted that the one-sided morphology of pc-scale jets was due to “Doppler favoritism”. Although, ideally, identical relativistic jets are expected to emerge symmetrically from the supermassive black hole, due to beaming and the preferred viewing angle of blazars, the jet moving towards us is boosted, whereas the counter-jet is deboosted. The emission from the jet coming towards us will be boosted by an amount equal to a power of the Doppler factor ($D$) defined as:
\begin{equation}
D=\frac{1}{\Gamma(1-\beta\cos\theta)},
\end{equation}     
where $\Gamma$ is the bulk Lorentz factor of the jet, $\theta$ is the viewing angle, and $\beta$ is the speed of the jet in units of speed of light. In the case of the receding jet the Doppler factor (substituting $\beta$ with $-\beta$) is:
\begin{equation}
D'=\frac{1}{\Gamma(1+\beta\cos\theta)}.
\end{equation}
For the boosted jet the monochromatic flux density is given by: 
\begin{equation}
S_\nu=\frac{L_\nu D^p}{4\pi{d_L^2}}(1+z)^{1+s},
\label{fluxdensity}
\end{equation}
and for the deboosted jet by:
\begin{equation}
S'_\nu=\frac{L_\nu D'^p}{4\pi{d_L^2}}(1+z)^{1+s},
\end{equation}
where $s$ the spectral index defined as $S\propto\nu^s$, $L_\nu$ the monochromatic intrinsic luminosity, $d_L$ is the luminosity distance, $z$ the redshift, and $p=2-s$ for the continuous jet case, and $p=3-s$ for the discrete. Throughout this work we have adopted the continuous jet case. The ratio of the two flux densities would be:
\begin{equation}
j=\frac{S_\nu}{S_\nu'}=\left(\frac{D}{D'}\right)^p=\left(\frac{1+\beta\cos\theta}{1-\beta\cos\theta}\right)^p.
\label{counter_ratio}
\end{equation}
If a source is closely aligned with our line of sight ($\theta$ is small, as is the case for blazars), then even for the mildly relativistic case the ratio will be approximately $\sim10^3$. For a more typical blazar jet with Lorentz factor of $\sim 15$, detection of the counter-jet is indeed very hard. For this reason, although the detection of pc-scale counter-jets is a complex problem that can be affected by several factors (see section \ref{constrains} for a detailed discussion) in the case of blazars Doppler favoritism is so strong that likely outweighs other difficulties.

In this work, we quantify, statistically, the expected distributions of $j$ and $S_\nu'$ produced by Doppler boosting alone, which set the {\em minimum} sensitivity requirements for the detection of blazar counter-jets by current and future observatories.

\subsection{Possible applications of blazar counter-jet detection}

Provided that other factors affecting the counter-jet flux density (such as free-free absorption from ionized material surrounding the black hole, see section \ref{constrains}) can be controlled, measurements of the jet-to-counter-jet ratio ($j$) can be used to constrain the jet viewing angle. Blazar jets are known to show superluminal motion \citep{Cohen1979,Pearson1981,Vermeulen1994,Lister2009-2,Lister2013}. Measured velocities of the radio components in the jets vary from little over the speed of light to $\sim 50c$ \citep{Lister2009-2,Lister2013}. 
Surveys such as the MOJAVE (Monitoring of Jets in Active galactic nuclei with VLBA Experiments, \citealp{Lister2005})\footnote{http://www.physics.purdue.edu/MOJAVE/} and the  Boston Blazar Research Group program\footnote{http://www.bu.edu/blazars/} \citep{Jorstad2001,Jorstad2005,Jorstad2006} have spearheaded great progress in reliably measuring apparent velocities for a large number of blazars.
The apparent superluminal motion is related to the true component speed ($\beta$) and the viewing angle ($\theta$) by:
\begin{equation}
\beta_{app}=\frac{\beta\sin\theta}{1-\beta\cos\theta}.
\label{app_velocity}
\end{equation}
Combining Eq. \ref{counter_ratio}, and Eq. \ref{app_velocity} we have:
\begin{equation}
\theta=\arctan\left(\frac{2\beta_{app}}{j^{1/p}-1}\right).
\end{equation}
Detection of the counter-jet  with the addition of a reliable measurement of the apparent velocity of the approaching jet may thus provide an independent estimate of the viewing angle, a key parameter in blazar modeling. Knowing the viewing angle in turn allows us to estimate the bulk Lorentz factor of the flow and the Doppler factor independently of the frequently used assumptions entering various Doppler factor estimation techniques (e.g. Variability Doppler factors \citealp{Lahteenmaki1999-III,Hovatta2009}). Such assumptions, like the equipartition of the energy density of the magnetic field and the radiation energy density, domination of synchrotron self-Compton emission at high energies, the proportionality of the synchrotron peak with luminosity etc., could be then tested in individual sources. Independent estimates of Doppler factors could also contribute to further constraining the “true” Doppler factor in individual sources, and aid in the identification of empirical correlations in blazar jets that have traditionally been obscured by relativistic effects and limitations in confidently estimating beaming properties (e.g., \citealp{Hovatta2010,Lister2011,Blinov2016,Blinov2016-II,Angelakis2016})

In addition to understanding the beaming properties of blazar jets, we would be able to investigate many other important aspects of blazar and AGN physics in general. For example, since the two jets are expected to be symmetric, the supermassive black hole will lie between the radio core and counter-core (e.g \citealp{Haga2013,Baczko2016}). Locating the supermassive black hole and of the broadline region with respect to the radio core would provide invaluable insight into localizing the emission regions in different wavelengths that could be complementary to other techniques (see, e.g., \citealp{Fuhrmann2014}, \citealp{Max-Moerbeck2014}, \citealp{Abdo2010-III}, \citealp{Blinov2015}).

\subsection{Observational detections of pc-scale counter-jets}

Counter-jets on pc-scales have been discussed in a number of one-sided jet sources, including Seyfert galaxies, radio galaxies, LINERS and blazars. These include 3C84 \citep{Vermeulen1994-counter}, Markarian 501 \citep{Giovannini2008}, PKS1510-089 \citep{Homan2002}, M87 (\citealp{Arp1967,Sparks1992,Kovalev2007}, Walker, Hardee, Davies, Ly, \& Junor in prep.\footnote{see also \url{http://www.aoc.nrao.edu/~cwalker/M87/}}). Nevertheless the vast majority of counter-jets remains undetected.

In the case of M87, an optical counter-jet was reported as early as 1967 \citep{Arp1967,Sparks1992}. More recently, radio counter-jet features have been detected in sub-pc scales in the inner jet of M87 (\citealp{Kovalev2007}, Walker, Hardee, Davies, Ly, \& Junor in prep.) However, the measured jet-to-counter-jet ratio (10-15) is inconsistent within the simple Doppler favoritism picture observed with the lack of evidence of fast motions within the inner 20 mas \citep{Kovalev2007}. The authors conclude that this may be due to either an intrinsic asymmetry  of the jet, or the lack of distinguishable moving features within a smooth fast flowing inner jet.

A pc-scale counter-jet was detected by \cite{Vermeulen1994-counter} in the case of 3C84. In this case the counter-jet featured an inverted spectrum, with a spectral index of 1.7, most naturally interpreted as a sign of free-free absorption by ionized, toroidal or disklike material around the black hole, which would not affect the approaching jet (see  also \citealp{Fujita2016}). Free-free absorption effects have also been identified in the spectra of counter-jet features in NGC 4261, Cen A, and Cyg A by \cite{Haga2013,Haga2015} and \cite{Boccardi2016}. When this is the case, caution needs to be exercised when interpreting the jet-to-counter-jet ratio, as a significant function of the flux reduction may be due to absorption compounding the effect of Doppler deboosting.

The precessing jet of 4C +12.50, which is a source whose host shows signs of recent merger, is another case of detected pc-scale counter-jet. In this very well-studied source, the detection of superluminal apparent speeds, jet and counter-jet, as well as bend morphology have allowed the fitting of a detailed model describing the jet precession \citep{Lister2003}.

In the case of PKS 1510-089 a kpc-scale counter feature has been detected by \cite{Odea1988} although more recently \cite{Homan2002} favor a jet bend interpretation of this structure. VLBI observations of the source have been conflicting on whether there  appears to be a pc-scale counter-jet towards the kpc-scale counter feature \citep{Bondi1996,Fey1997,Kellerman1998,Homan2001,Jorstad2001,Wardle2005,Homan2002}.

Finally, in the case of Markarian 501, \cite{Giroletti2004,Giroletti2008} used the non-detection of a counter-jet in high dynamic range observations to constrain the jet geometry.

\subsection{This work}
In this work, we use re-optimized population models (of the same class as \citealp{Liodakis2015}, hereafter Paper I) using the $\beta_{app}$ and redshift distributions from the MOJAVE survey \citep{Lister2005}  in order to describe statistically the underlying distribution of bulk Lorentz factors $\Gamma$, unbeamed luminosities ($L_\nu$), viewing angles and redshifts for the blazar population. We then use these models to produce the distribution of the jet to counter-jet ratio, and the distribution of the counter-jet flux-density in different radio frequencies. We aim at predicting the minimum sensitivity required from future telescopes, and upgrades of the existing arrays in order to  detect counter-jets from blazars. 

This paper is organized as follows: In section \ref{newmodel} we present the modifications to the population model from Paper I; in section \ref{det_coun_jet} we present the models prediction for the jet-to-counter-jet ratio and the flux density of the counter-jet in different frequencies; in section \ref{candidates} we present possible counter-jet detection candidates; and in section \ref{discussion_and_conclusion} we discuss the conclusions derived from this work.

The cosmology we have adopted throughout this work is $H_0=71$ ${\rm km \, s^{-1} \, Mpc^{-1}}$, $\Omega_m=0.27$ and $\Omega_\Lambda=1-\Omega_m$ \citep{Komatsu2009}.

\section{Population Model}\label{newmodel}  

In Paper I, separating BL Lacs and FSRQs, we created population models i.e. a bulk Lorentz factor and unbeamed luminosity distributions that can adequately reproduce the observed apparent velocity of the radio components and redshift distributions from the statistically complete, flux-limited sample of the MOJAVE survey \citep{Lister2005}. 

We determined the fraction of sources at each redshift, from $z=0$ to $z=1.4$ for the BL Lacs and from $z=0$ to $z=2.5$ for the FSRQs, assuming pure luminosity evolution.  We assumed random uniformly distributed values for the viewing angle ($\cos\theta$ distributed from 0 to 1) and power-law distributions for the Lorentz factor and the unbeamed luminosity. From these distributions we drew simulated source samples, consisting of all the sources with flux density above 1.5 Jy, which is the flux-limit of the MOJAVE survey. The number of simulated source draws was such that the final sample consisted of $\sim 10^3$ sources. We then used the Kolmogorov-Smirnov test (K-S test) to determine the consistency of simulated and observed $\beta_{app}$ and $z$ distributions in order to select ranges of acceptable model parameters. The parameters we optimized are the power-law indices for the unbeamed luminosity distribution (A) and for the Lorentz factor distribution ($\alpha$), and the evolution parameter ($\tau$) for the FSRQs. For the BL Lacs we only optimized for the power law indices (A,$\alpha$) due to the lack of evolution in the BL Lac luminosity function (\citealp{Ajello2014}, Paper I). For a more detail explanation of the models, optimization techniques as well as applications see Paper I, and \cite{Liodakis2015-II}.

Throughout Paper I we had assumed a single spectral index for each of the BL Lac and FSRQ classes. This assumption was made for simplicity reasons, although in reality not only the spectral index is different for each source, but it is also time-variable \citep{Angelakis2012}. Since our results might be sensitive to the spectral index of each source, we need a more accurate representation than simply assuming a single spectral index for each population. For this reason, we have included a spectral index distribution for each of the blazar classes. We have incorporated a known spectral index distribution measured from observations independently (\citealp{Hovatta2014}, MOJAVE survey), and then re-optimized the same parameters (A, $\alpha$, $\tau$) for the models as in Paper I.

\subsection{Spectral index distribution}\label{spindexdistr}

\begin{figure}
\resizebox{\hsize}{!}{\includegraphics[scale=1]{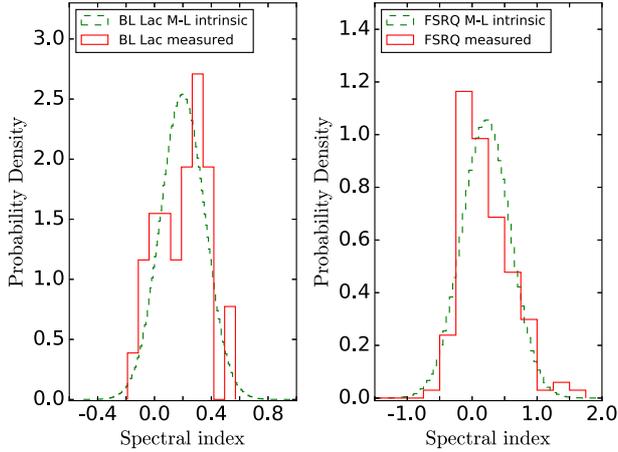}}
 \caption{Distribution of the core spectral index. Left panel: BL Lacs; right panel: FSRQs. The red solid line is the histogram of spectral indices \citep{Hovatta2014}, whereas the green dashed line the derived distribution of the population spectral index having taken into account the error in the measurements. }
 \label{plt_spec_index_bl_qs}
 \end{figure}

For our spectral index distributions we used spectral index data from the MOJAVE syrvey \citep{Hovatta2014}. We caution that due to the flux-density limit of the MOJAVE survey, only a small number of BL Lacs is included in the sample in comparison to FSRQs. It is possible that the BL Lac sample is a somewhat biased subsample of the class since it only includes the most luminous BL Lacs at 15 GHz. Since the radio core of the jet is dominating the emission \citep{Lister2005,Cooper2007}, we use only the core spectral indices. Assuming that the spectral index distribution of the population is normally distributed, we estimate the mean and standard deviation of each population taking into account the errors of each measurement. In order to properly account for errors, we follow the analysis described in \cite{Venters2007}. Since all the reported values have the same error, the mean of the distributions is given by:
\begin{equation}
s_{tr}=\frac{\sum^N_{j=1}s_j}{N},
\label{spid_mean}
\end{equation}
where $s_{tr}$ is the true mean of the distribution, and $s_j$ are the observed data. The standard deviation would be:
\begin{equation} 
\sigma_{tr}=\sqrt{\frac{\sum^N_{j=1}(s_j-s_{tr})^2}{N}-\sigma_j^2},
\label{spid_std}
\end{equation}
where $\sigma_{tr}$ is the true standard deviation of the distribution, and $\sigma_j$ is the reported error of each measurement. Using this method we calculated that the mean for the BL Lacs is $s_{tr}=0.19$ with standard deviation $\sigma_{tr}=0.15$ and for the FSRQs $s_{tr}=0.21$ with standard deviation $\sigma_{tr}=0.37$. Figure \ref{plt_spec_index_bl_qs} shows the distribution of the spectral index for the BL Lac and FSRQ populations.
Simply using data from the literature for this calculation without taking errors into account will overestimate the spread of the distribution.

\subsection{A re-optimized model}\label{reopt-model}

\begin{figure}
\resizebox{\hsize}{!}{\includegraphics[scale=1]{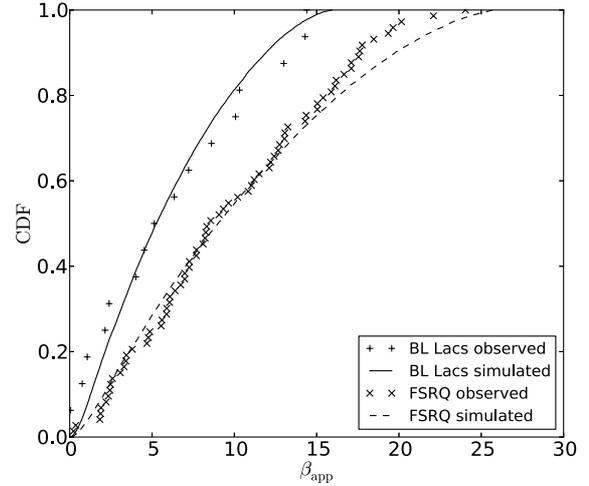} }
 \caption{Cumulative distribution function of the apparent velocity for the BL Lacs and the FSRQs.}
 \label{plt_bobs_all}
 \end{figure}
\begin{figure}
\resizebox{\hsize}{!}{\includegraphics[scale=1]{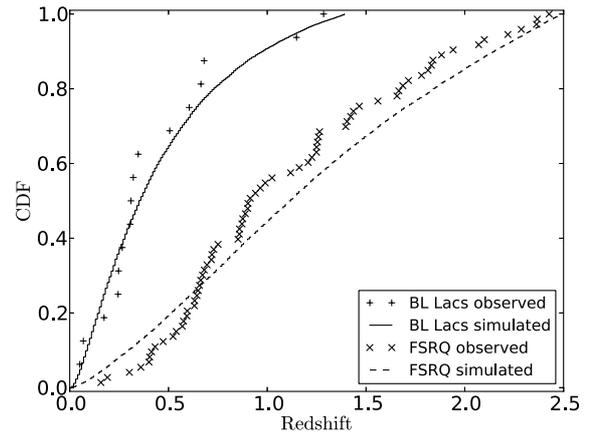} }
 \caption{Cumulative distribution function of the redshift for the BL Lacs and the FSRQs.}
 \label{plt_redshift_all}
 \end{figure}
In order to optimize our models, we use Monte-Carlo sampling as follows:

\begin{flushleft}
For each source we:
\begin{itemize}
\item {draw a value for the viewing angle from a uniform distribution ($\cos\theta$ uniformly distributed \footnote{uniformity of the cosine corresponds to the  assumption of equal a priori probability to observe a given jet aligned with any line of sight, see e.g. \cite{Tassis2007,Tassis2009}} between [0,1]),}
\item {draw a value for the bulk Lorentz factor and an unbeamed luminosity from a powerlaw distribution \citep{Padovanni1992,Lister1997},}
\item {draw a value for the spectral index from a normal distribution with the mean and standard deviation calculated in section \ref{spindexdistr} for each population.}
\end{itemize}
\end{flushleft}
We calculate a flux-density using Eq. \ref{fluxdensity} and apply the 1.5 Jy flux-limit. When we have our final simulated sample, we construct the cumulative distribution functions for the apparent velocity and the redshift and use the Kolmogorov-Smirnov test in order to test if our model is acceptable. Following the above procedure we generate $10^4$ samples, each time drawing a random value for the parameters uniformly. We consider a model acceptable if the joint K-S test yields a $\geq 5\%$ probability of consistency between observed and simulated samples which is our null hypothesis.

The ``best-fit'' values of the model parameters as well as ranges of parameters that yield acceptable models are summarized in Table \ref{tab:Parameters}. We note that due to the lack of evolution in BL Lacs, the FSRQ population as a whole turns out to be considerably more luminous in our models, as is known to be the case. In addition, although all viewing angles start out equally probable, the application of a flux limit imparts a strong preference for viewing angles $\lesssim 1/\Gamma$ (see also paper I).

\begin{table*}
\begin{minipage}{180mm}
\setlength{\tabcolsep}{11pt}
\centering
  \caption{Optimal parameter values for our population models. The asymmetrical uncertainties indicate the range within which a parameter can produce ``acceptable models". The values quoted for the spectral index are the mean and standard deviation of the population and are derived  from observations independently (not through Monte-Carlo fitting), under the assumption that the spectral index is independent from other source properties.}
  \label{tab:Parameters}
\begin{tabular}{@{}cccccc@{}}
 \hline
   &   & BL Lacs (This work) &  FSRQs (This work) & BL Lacs (Paper I) & FSRQs (Paper I) \\
  \hline
  Lorentz factor &$\alpha$ & $0.50^{+1.04}_{-0.5}$ & $0.68^{+1.29}_{-0.68}$ & 0.73$^{+0.41}_{-1.46}$ & 0.57$^{+0.12}_{-0.50}$\\
  & & \\
  \hline
 Luminosity &A & $2.14^{+1.00}_{-0.76}$ & $2.42^{+1.58}_{-0.52}$& 2.25$^{+0.68}_{-0.78}$ & 2.60$^{+0.185}_{-0.245}$   \\
  & & \\
  &$\tau(1/H_0)$ & - & $\tau=0.29^{+0.05}_{-0.09}$ & - & 0.26$^{+0.068}_{-0.003}$\\
  & & \\
  \hline
Spectral index & s & $0.19\pm0.15$ & $0.21\pm 0.37$ & - & -\\
\hline
\end{tabular}
\end{minipage}
\end{table*}
For the apparent velocity the probability of consistency between observed and simulated samples using our best-fit models is $93.5\%$ for the BL Lacs, and $81.3\%$ for the FSRQs (Fig. \ref{plt_bobs_all}). For the redshift, the probability values are $70.4\%$ for the BL Lacs and $29.1\%$ for the FSRQs (Fig. \ref{plt_redshift_all}). Comparing this version of the model with that of Paper I, we see that the addition of the spectral index distribution resulted in small changes in the parameters and the probabilities of consistency. There is a slight flattening of the powerlaw distributions of the  intrinsic luminosity for both populations, and for the Lorentz factors for the BL Lacs, whereas we see a similar steepening of the Lorentz factor distribution for the FSRQs, all within the previously reported limits of parameter values that produced acceptable models. Also the evolution parameter value for the FSRQs has a higher value, which translates into weaker evolution of the intrinsic luminosity with redshift. Moreover, the probability values of consistency between observed and simulated samples have significantly increased for all distributions except the BL Lac apparent velocity which remained at the previously high levels ($\sim 93.5\%$). The probability of consistency for redshift of the BL Lacs rose from $54.1\%$ to $70.4\%$. In the case of the FSRQs the probability of consistency for apparent velocities rose from $49.3\%$ to $81.3\%$, and for the redshift from  $8.4\%$ to $29.1\%$.  

An important difference between this model and other population studies of blazars is that we make no attempt to reconstruct the blazar luminosity function (we optimize a slope but not an amplitude of the luminosity distribution); this simplification allows us to forego using the flux density (beyond the application of a flux limit) as an observable, thus being much less sensitive to the effect of variability.

The sample we used for the optimization was the sample defined in Paper I. Starting from the MOJAVE sample, we had excluded all the sources with no apparent velocity and redshift measurements. The implicit assumption associated with this choice is that these sources will follow the distributions of the sources for which measurements exist. To test the validity of this assumption, we created two test-samples (one for each population) and performed the optimization anew. In the test-samples we included all the sources with no measurements either in apparent velocity or redshift now under the assumption that sources without measurements have redshifts and apparent velocities typically higher and lower, respectively, compared to sources for which $z$ \& $\beta_{app}$ measurements exist. If the apparent velocity was not known, we assigned that source a random value (uniformly) from 0 up to the minimum known value in the population. If the redshift was not known (occurred only in the BL Lac population), we assigned that source a random value, again uniformly,  from the maximum known value in the population up to 2. Given the slope of the already known redshift distribution in our sample (Fig. \ref{plt_redshift_all}), it is unlikely that the BL Lacs will extend to redshifts much higher than that.

In Paper I we had also excluded two sources for being outliers. One for having abnormally high mean apparent velocity ($\beta_{app}\sim 50$) and one due to high redshift ($z=3.396$). In the updated online data from \cite{Lister2013} the former has now a mean apparent velocity of $\sim 22$ which is within the limits of the original sample and it is no longer considered an outlier. Both of the these sources were included in the test-samples.

We were not able to produce an acceptable model (joint K-S test $\geq$5\%) for either the BL Lac or the FSRQ populations. This fact can be interpreted in two ways: Either the models we adopt cannot adequately describe blazars, or the assumptions used to construct the test-samples were unrealistic. Given that our models are consistent with generally used assumptions regarding the blazar population, the latter possibility is preferred.

It is in principle possible that some other treatment of unknown z and $\beta_{app}$ than the one we adopted, which however still places these quantities outside observed ranges, can produce fits. However, it is more likely that the sources without apparent velocity or redshift do not lay outside the limits of the Paper I sample but reside within. Any lack of estimates for the missing quantities is most likely due to other observational constrains such as weak emission lines in the case of $z$. If we consider the test-sample ``fit'' as indicative, $\beta_{app}$ outside the observed range that may exist in the censored sources will drive the model towards a steeper distribution of $\Gamma$ and a flatter distribution of luminosities.

Additionally, we performed the optimization once more for the FSRQ sample, but this time we included only the redshift outlier. The best-fit model gives a $\sim$82\% probability of consistency for the apparent velocity and $\sim$6\% for the redshift, producing a marginally acceptable 5\% joint probability. The best-fit parameters for this model are $\alpha=1.57$, $A=3.41$, and $\tau=0.24$. All the parameter values are within the limits of the parameter space that produces acceptable models (section \ref{reopt-model}). There is a steepening of the power-law indices for the Lorentz factor distribution and luminosity function, as well as change in $\tau$, supporting stronger luminosity evolution. Although the probability of consistency for the apparent velocity is the same, it is not true for the redshift. There is a significant drop in the probability value, which falls even lower than the original model (Paper I).

We have limited the parameter space for the Lorentz factor power law index ($\alpha$) to positive values. Negative values would imply a larger number of sources with high than with low $\Gamma$. Judging from the $\beta_{app}$ distribution \citep{Lister2009-2,Lister2013} as well as Lorentz factor distributions from variability studies \citep{Hovatta2009,Liodakis2016} such a scenario is unlikely. However, if we do not limit ourselves to positive values, the best-fit values do not change, but the range of parameters that produce acceptable models becomes $\alpha$--[-1.55,1.54], A--[1.38,3.14] for the BL Lacs and $\alpha$--[-0.14,1.97], A--[1.88,4.0], $\tau$--[0.19,0.34] for the FSRQs

In theory, there is the possibility of a dependence of the intrinsic luminosity to the bulk Lorentz factor which would further complicate the situation. However the ability of the model to naturally reproduce the observed 15 GHz flux-density distributions of the two populations (Paper I) gives us confidence that any effect of such a dependence would not alter our results dramatically.

\section{Detectability of the counter-jet}\label{det_coun_jet}

Using the optimized models obtained with the procedure described above, we will derive the ratio of the approaching and the receding jet flux densities, and the expected distribution of counter-jet flux-densities at different frequencies, assuming that the two jets are intrinsically identical. We will then attempt to establish the sensitivity required for future experiments to achieve detection of the counter-jets.

\subsection{Jet-to-counter-jet ratio}
\begin{figure}
\resizebox{\hsize}{!}{\includegraphics[scale=1]{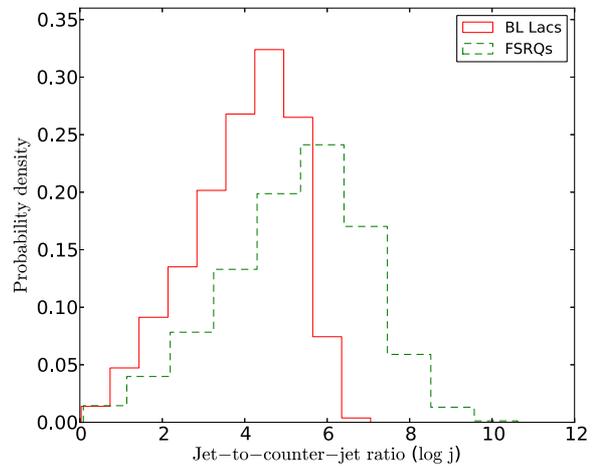} }
 \caption{Distribution of the logarithm of the jet-to-counter-jet ratio  for the BL Lacs (solid red) and the FSRQs (dashed green) for our best-fit models.}
 \label{plt_ratio_all}
 \end{figure}
\begin{figure}
\resizebox{\hsize}{!}{\includegraphics[scale=1]{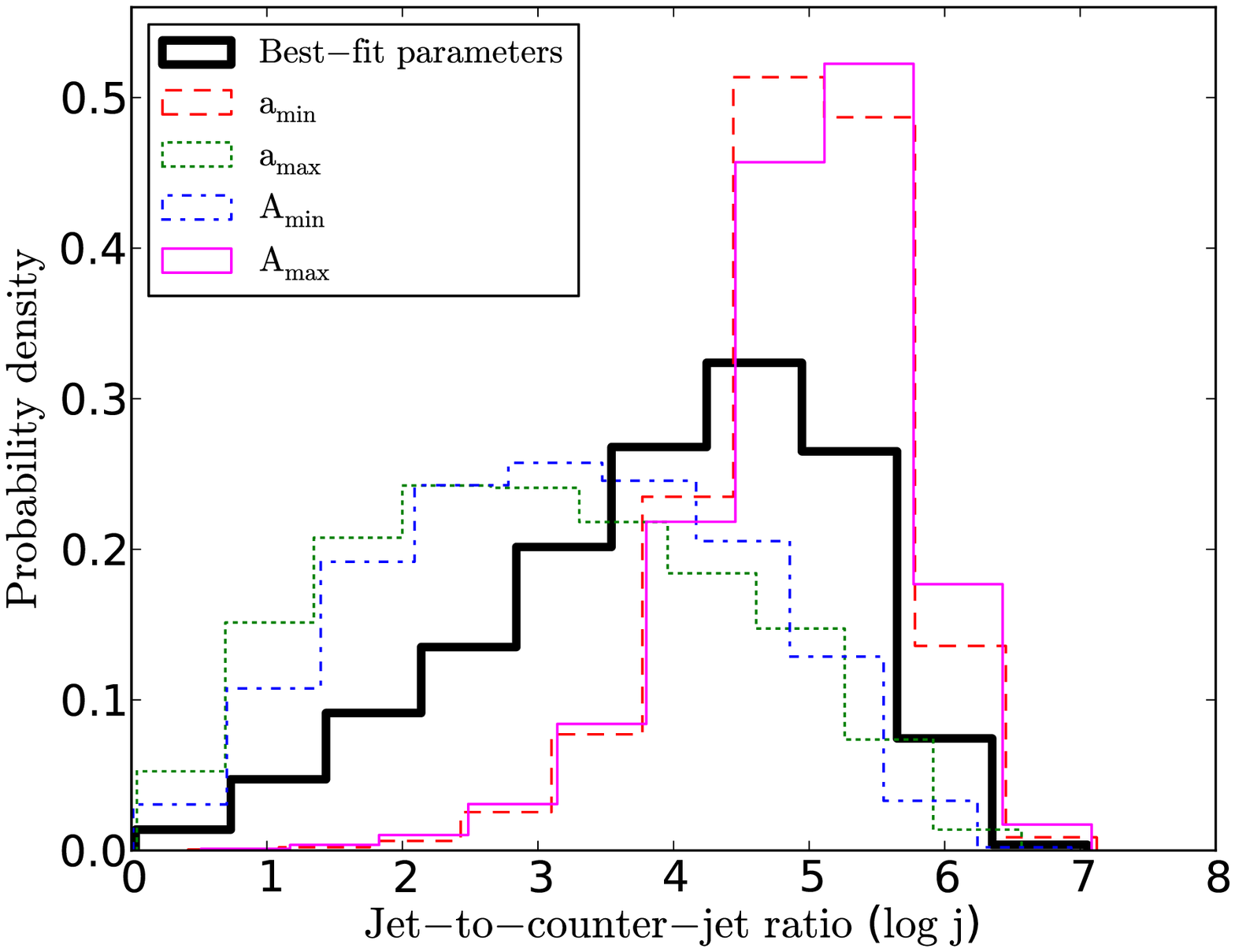} }
 \caption{Distribution of the logarithm of the jet-to-counter-jet ratio optimal model (black solid line) and the models at the extrema of the parameter space for each parameter for the BL Lacs.}
 \label{plt_ratio_bl_limits}
 \end{figure}
\begin{figure}
\resizebox{\hsize}{!}{\includegraphics[scale=1]{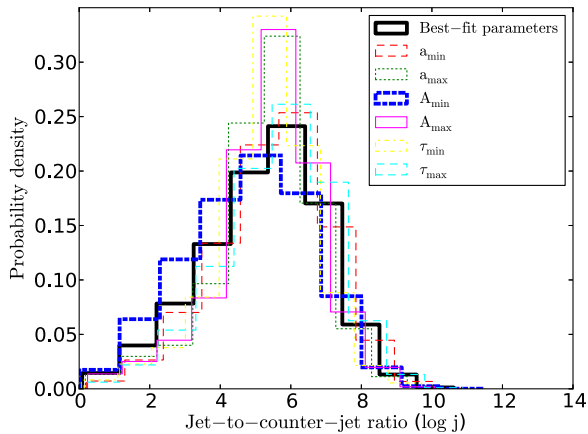} }
 \caption{Distribution of the logarithm of the jet-to-counter-jet ratio optimal model (black solid line) and the models at the extrema of the parameter space for each parameter for the FSRQs.}
 \label{plt_ratio_qs_limits}
 \end{figure} 
In order to estimate the flux-density of the counter-jet, we use the values for the velocity of jet in units of speed of light ($\beta$), as well as the spectral index of each simulated source, and calculate the jet-to-counter-jet ratio using Eq.\ref{counter_ratio}. Figure \ref{plt_ratio_all} shows the distribution of the logarithm of the jet-to-counter-jet ratio ($\log j$) for the BL Lac and the FSRQ population. The minimum value for the $\log j$ is  $0.03$ for the BL Lacs and $0.08$ for the FSRQs, whereas the maximum is $\sim 7$ and $\sim 10.6$ for the BL Lacs and FSRQs respectively. It is clear from Fig. \ref{plt_ratio_all} that the BL Lacs have smaller values of $\log j$ than the FSRQs with a sharper distribution peaked between $4.2$ and $4.9$ , and with a mean of $\sim 3.9$. The FSRQ population has a wider distribution with a peak between $5.3$ and $6.4$ ,  mean at $\sim 5.2$, and a tail extending up to almost four orders of magnitude higher than the BL Lacs.

Figures \ref{plt_ratio_bl_limits} and \ref{plt_ratio_qs_limits} show the distribution of the jet-to-counter-jet ratio for different models for the BL Lacs and FSRQs respectively. Each distribution corresponds to the resulting ``acceptable'' model at the extreme value of one of the parameters. Any shift in the distribution of the jet-to-counter-jet ratio will result in  changes of the expected counter-jet flux density distribution and thus the number of counter-jets that could possibly be detected. In the case of the FSRQs (Fig. \ref{plt_ratio_qs_limits} ) we see little shift in the jet-to-counter-jet ratio distribution for the different models. The percentage of sources above certain flux-density values (see section \ref{Det_limits}) only changes by a few percent either rising or falling depending on the model. Thus our results will not be affected significantly within our parameter space. For the BL Lacs the situation is somewhat different (Fig. \ref{plt_ratio_bl_limits}). There is a considerable shift in the distributions for the different models. Such a shift could result in changes of the fraction of detectable counter-jets up to 20-30 percentage points, depending on the parameter. The reason for this difference is that the FSRQ sample on which our model optimization is based is considerably larger than the BL Lac one, and as a result the FSRQ model parameters are better constrained.

\subsection{Detectability limits}\label{Det_limits}
\begin{figure}
\resizebox{\hsize}{!}{\includegraphics[scale=1]{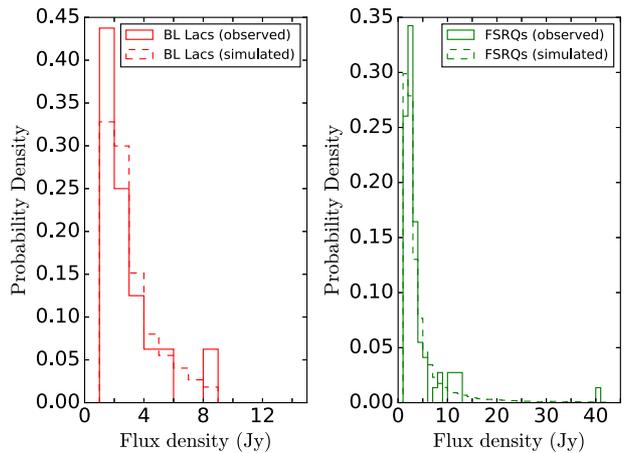}}
 \caption{Distribution of the observed and simulated flux densities for both BL Lacs (left panel) and FSRQs (right panel). Solid red is for the observed BL Lacs, dashed red for the simulated BL Lacs, solid green is for the observed FSRQs, dashed green for the simulated FSRQs.}
\label{plt_flux_index_bl_qs}
 \end{figure}

\begin{figure}
\resizebox{\hsize}{!}{\includegraphics[scale=1]{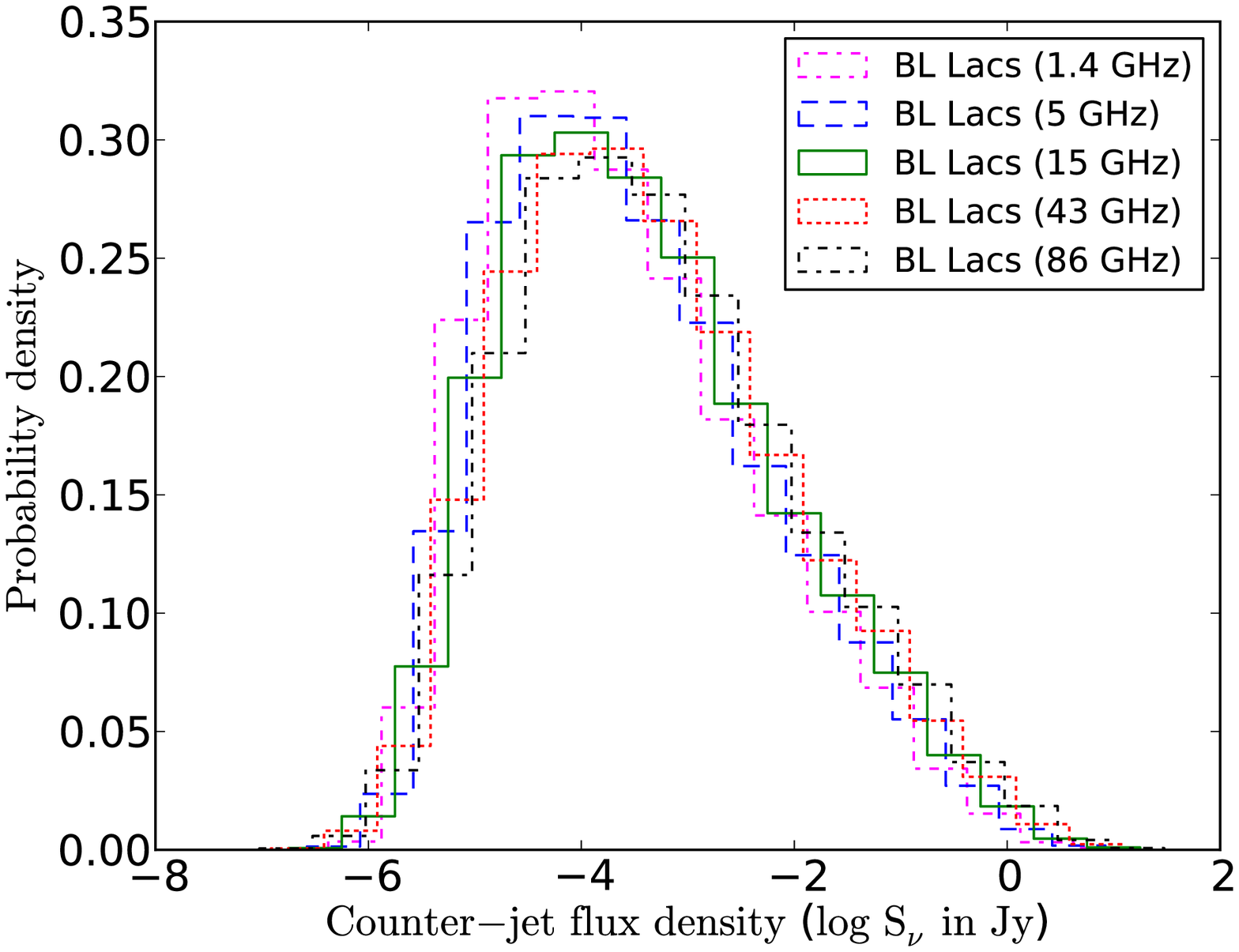} }
 \caption{Distribution of the logarithm of the counter-jet flux density at different frequencies for the BL Lac population. Dash-dot magenta is for the 1.4 GHz, dashed blue is for the 5 GHz, solid green for the 15 GHz, dotted red for the  43 GHz, and dash-dot black for the 86 GHz.}
 \label{plt_counter_bl}
 \end{figure}
\begin{figure}
\resizebox{\hsize}{!}{\includegraphics[scale=1]{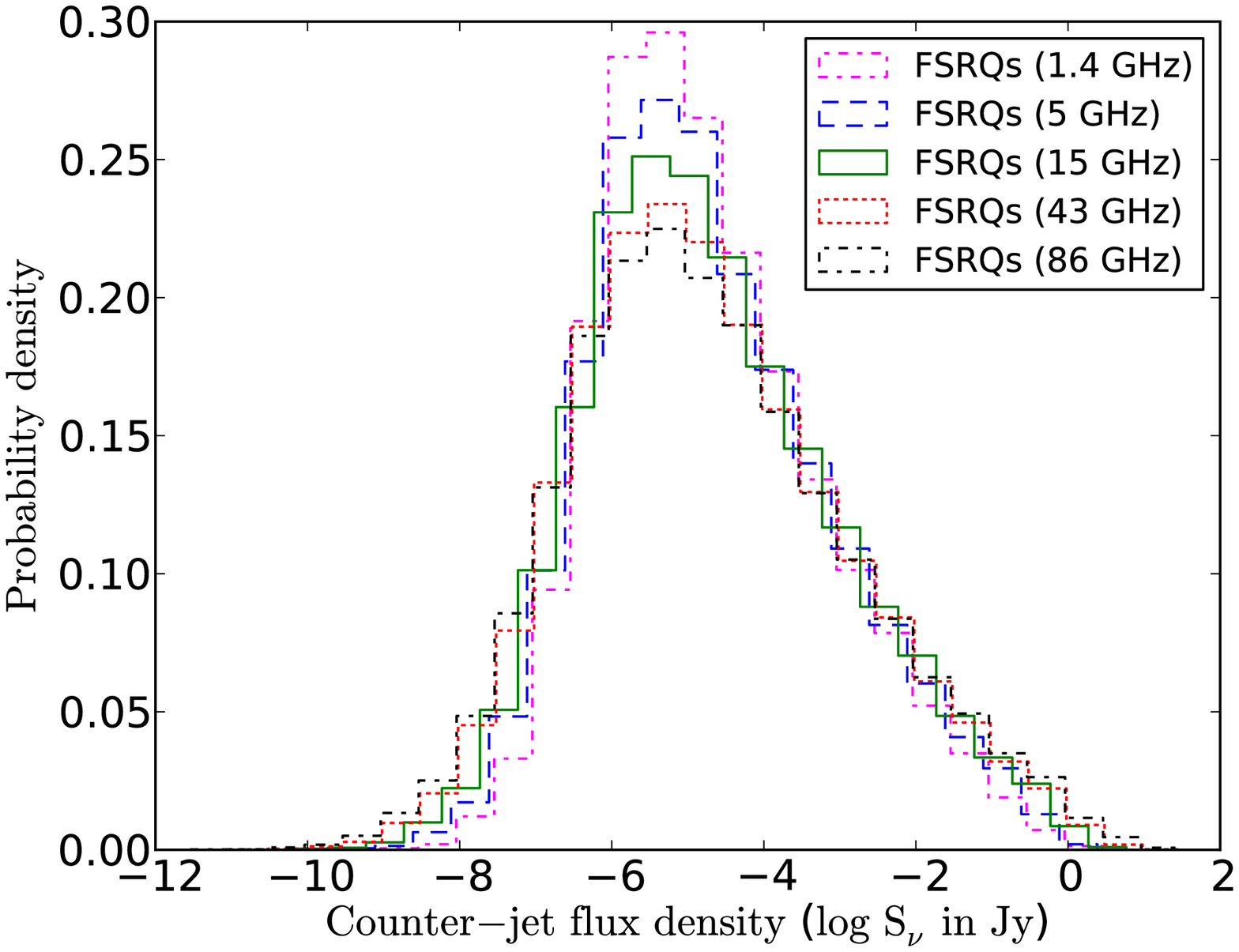} }
 \caption{Distribution of the logarithm of the counter-jet flux density at different frequencies for the FSRQ population. Dash-dot magenta is for the 1.4 GHz, dashed blue is for the 5 GHz, solid green for the 15 GHz, dotted red for the  43 GHz, and dash-dot black for the 86 GHz.}
 \label{plt_counter_qs}
 \end{figure}
\begin{figure}
\resizebox{\hsize}{!}{\includegraphics[scale=1]{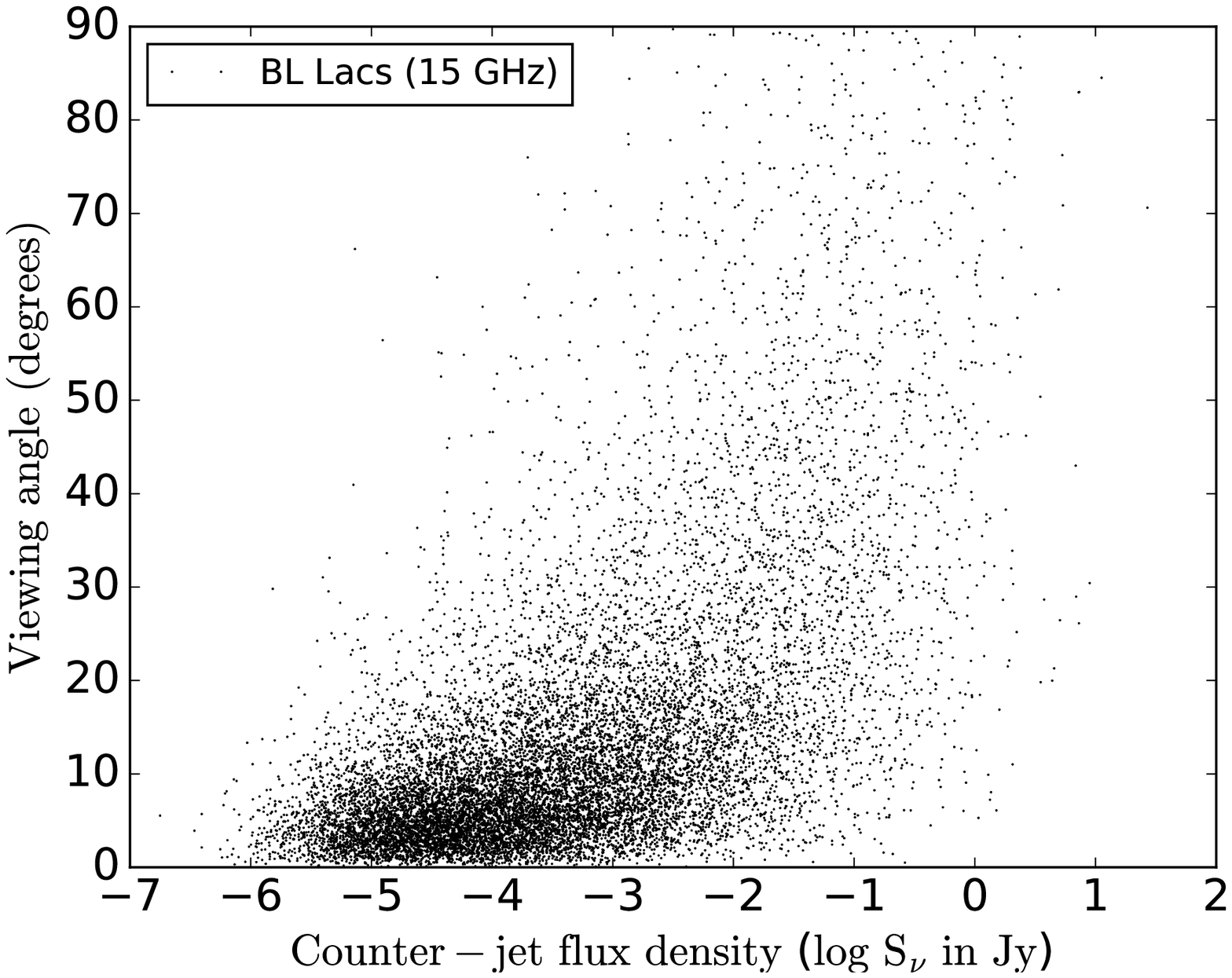}  }
 \caption{Viewing angle plotted against the counter-jet flux density at 15 GHz for the BL Lac population.}
 \label{plt_thcounter_bl}
 \end{figure}
\begin{figure}
\resizebox{\hsize}{!}{\includegraphics[scale=1]{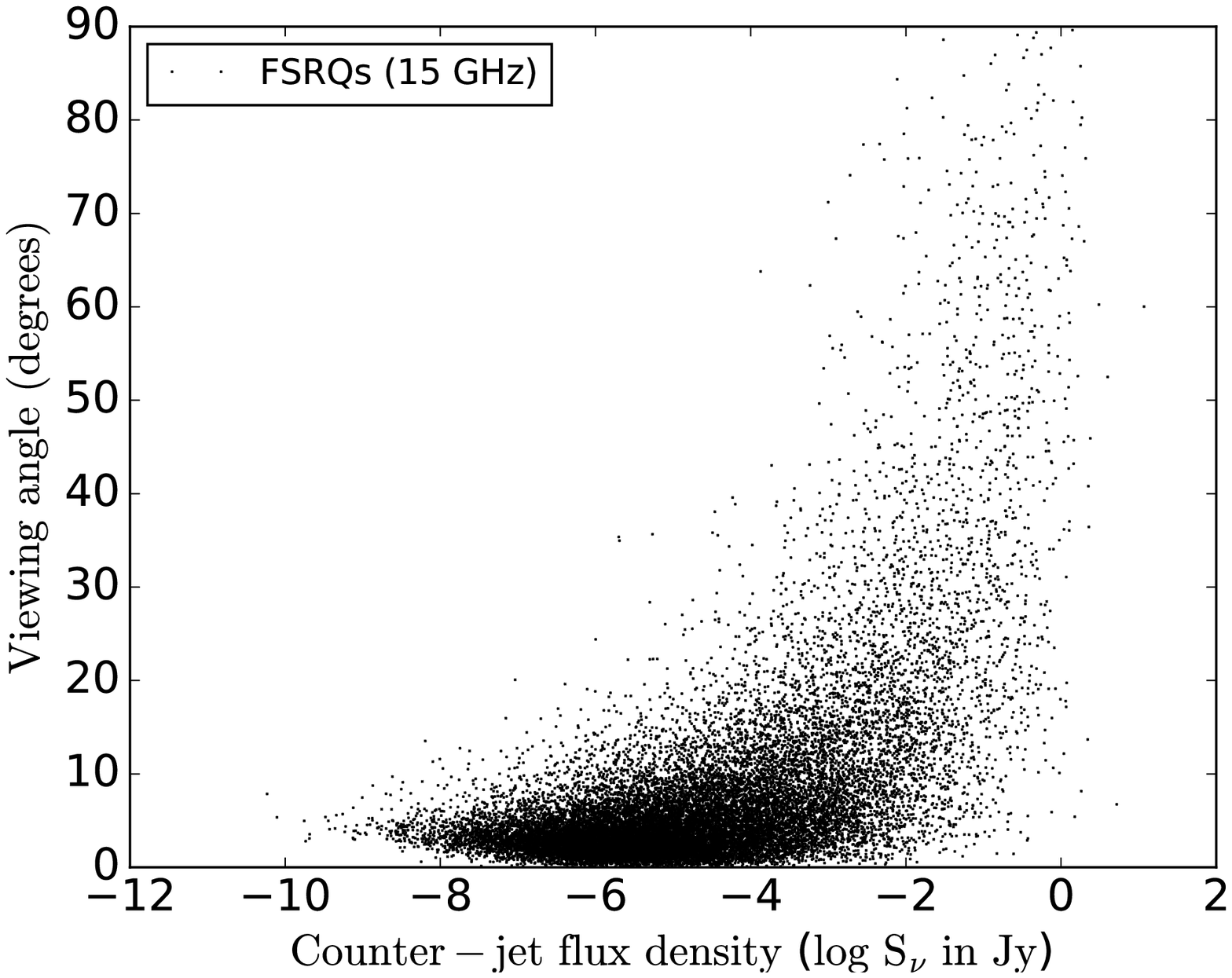}  }
 \caption{Viewing angle plotted against the counter-jet flux density at 15 GHz for the BL Lac population.}
 \label{plt_thcounter_qs}
 \end{figure}

Our models for the blazar population do not use the flux density as an observable for the model optimization (other than as a flux limit at 15 GHz). We can however compute the predicted flux-density distributions of both jet and counter-jet at 15 GHz using our model distributions for the luminosity, Lorentz factor, redshift, and spectral index for each population, assuming random viewing angles. Figure \ref{plt_flux_index_bl_qs} shows the 15 GHz flux density distributions for the jets of BL Lacs and FSRQs overplotted with observations. To compute the flux-density of the primary jet at frequencies other than 15 GHz, we use the spectral index for each source ($> 10^4$ sources for each class) and equation:
\begin{equation}
S_n=\left(\frac{\nu_n}{\nu_{m}}\right)^sS_{m},
\label{flux_conv}
\end{equation}
where n and m indicate the frequency. Then, we use the jet-to-counter-jet ratio (Eq. \ref{counter_ratio}) in order to calculate the counter-jet flux density at different frequencies. For the present analysis, we examined  four other frequencies, 1.4, 5, 43 and 86 GHz. We chose these four frequencies because they are the most widely used and/or will be used in radio interferometric experiments (e.g ALMA, SKA, VLBI) apart from 15 GHz. Figures \ref{plt_counter_bl} and \ref{plt_counter_qs} show the distribution of the counter-jet flux density for different frequencies, for the BL Lacs and FSRQs respectively. The FSRQ population has a more centrally concentrated distribution centered around lower values than that of the BL Lacs. The maximum for both populations is $\sim 1$ Jy whereas the minimum for the BL Lacs is $\sim 10^{-6.7}$ and for the FSRQs is $\sim 10^{-10.2}$Jy. The peak of the BL Lac distribution is around $\sim 10^{-4} $Jy whereas for the FSRQs the peak is approximately an order of magnitude lower $\sim 10^{-5} $Jy.

The counter-jet flux-densities can reach values as low as a few $\mathrm{\mu}$Jy and even nJy. However we find that $5\%$ of the BL Lac counter-jets have a flux-density higher than 100mJy, $15\%$ are higher than 10 mJy, and $32\%$ have higher flux-density than 1 mJy. For the FSRQs $8\%$ have a flux-density higher than 10mJy, $17\%$ are higher than 1 mJy, and $32\%$ are higher than 0.1 mJy. Although these percentages were calculated using the 15 GHz counter-jet flux density, we find no significant change in the values quoted above for the 1.4, 5, 43, and 86 GHz flux-densities. The percentages for the 1.4 and 5 GHz flux-densities are slightly shifted to higher values whereas for the 43 and 86 GHz are shifted to lower values for the same flux-density limits set above for both populations.  

In addition, we explore the connection between the viewing angle and the predicted counter-jet flux density.  Figures \ref{plt_thcounter_bl} and \ref{plt_thcounter_qs} show the viewing angle (in degrees) plotted against the counter-jet flux density (in Jy) for the BL Lacs and FSRQs respectively. The FSRQ population seem to be more concentrated on the $\theta$-$S_\nu$ plane than the BL Lacs. This can be attributed to the FSRQs having, on average,  smaller viewing angles and faster jets. For the ``typical" blazar case with viewing angle $\leq 15^o$, 68.5\% of the BL Lacs and 86.2\% of the FSRQs have counter-jet flux density $\leq 1$ mJy. Although there are outliers, for the vast majority of sources in both populations there seems to be a trend were smaller viewing angle translates to fainter counter-jet. 

There are also cases, especially in the BL Lac population, were sources with viewing angle larger than $45^o$ have counter-jet flux densities lower than 1 mJy, suggesting that the velocity of the jet ($\beta$) is high enough to conceal the receding jet even with a large viewing angle. Within the current unification scheme, BL Lacs objects are the beamed counterparts of FR I galaxies, implying that the only difference between the two is the orientation of the jet to our line of sight. BL Lac objects have small viewing angles, typically $\leq 15^o$, whereas FR I galaxies have viewing angles $\geq 30^o$ \citep{Ghisellini1993}. In \cite{Giommi2012} the authors find through Monte-Carlo simulation that $5\%$ of the radio selected and  $15\%$ of the X-ray selected blazars are misclassified as radio galaxies.  The BL Lac objects, in our simulated sample, with large viewing angles and one-sided jet morphology could resemble FR I galaxies. These sources, rare as they may be, could account for cases were a source is classified as a radio galaxy even though it is a blazar. Difficulties in the separation of the two classes are only in support of the unification view of FR I galaxies as the parent population of BL Lac objects.

\section{Counter Jet Detection Candidates}\label{candidates}
\begin{figure}
\resizebox{\hsize}{!}{\includegraphics[scale=1]{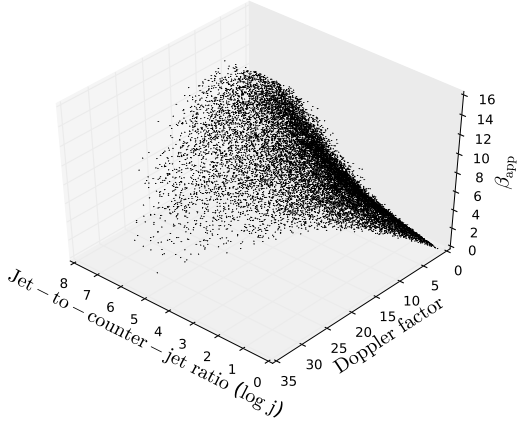} }
 \caption{Jet-to-counter-jet ratio versus the apparent velocity and Doppler factor for the BL Lacs.}
 \label{plt_3d-bllac}
 \end{figure}
\begin{figure}
\resizebox{\hsize}{!}{\includegraphics[scale=1]{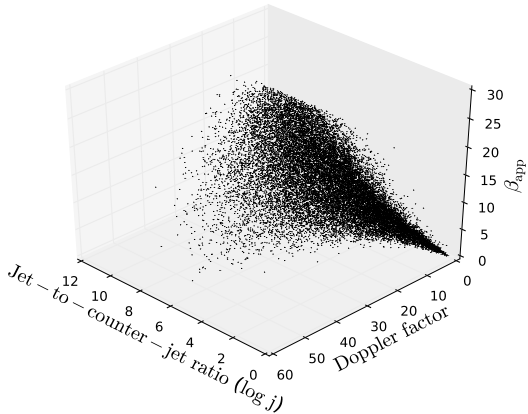}  }
 \caption{Jet-to-counter-jet ratio versus the apparent velocity and Doppler factor for the FSRQs.}
 \label{plt_3d-qso}
 \end{figure}
In order to identify the observables that could indicate a potential counter-jet detection candidate, we explored the relation between the jet-to-counter-jet ratio, the apparent velocity and the Doppler factor. Figures \ref{plt_3d-bllac} and \ref{plt_3d-qso} show the 3-D representation of the above observables for the BL Lacs and FSRQs respectively. Our simulations show that sources with both low apparent velocity ($\lesssim 3$) and low Doppler factor ($\lesssim 5$) have typically jet-to-counter-jet ratios $\lesssim 10^{4}$ which makes them prime detection candidates. Although we found no trend favoring specific redshift ranges, it is obvious that sources at lower redshifts are favored in terms of spatial resolution.

For specific sources if both $\beta_{app}$ and Doppler factor are known, then $\Gamma$ (and therefore $\beta$) and $\theta$ can be derived. If the spectral index (s) is also known, then the jet-to-counter-jet ratio can be calculated. Due to the MOJAVE survey \citep{Lister2005}, accurate estimates for the apparent velocity exist for more than 100 sources. On the other hand, not all Doppler factor estimates are reliable. We have shown in \cite{Liodakis2015-II} that the variability Doppler factors \citep{Lahteenmaki1999-III,Hovatta2009} can adequately describe both the BL Lac and FSRQs populations. An error analysis showed that the variability estimates are subject to systematic errors due to the cadence of observations. However, for low Doppler factors ($\lesssim 5$) the uncertainty of the estimates is dominated by statistical uncertainties, on average 30\%  \citep{Liodakis2015-II}. Thus we took to the literature to identify potential candidates that meet the above criteria. We found five such sources, for which we used data from \cite{Lister2013} for the apparent velocity, \cite{Hovatta2009} and \cite{Liodakis2016} for the Doppler factors,  \cite{Hovatta2014} for the core spectral index, and \cite{Lister2005,Lister2016} for the mean 15~GHz core flux density, in order to calculate the jet-to-counter-jet ratio and the expected counter-jet flux density at 15~GHz. All five sources as well as the relevant parameters are summarized in Table \ref{tab:candidates}.
\begin{table*}
\begin{minipage}{170mm}
\setlength{\tabcolsep}{11pt}
\centering
  \caption{Counter-jet detection candidates that meet the criteria set in section \ref{candidates}. Column (1): Name; (2): Redshift; (3): Doppler factor \citep{Hovatta2009,Liodakis2016}; (4): mean apparent velocity \citep{Lister2013}; (5): viewing angle (degrees); (6): core spectral index \citep{Hovatta2014}; (7): logarithm of the jet-to-counter-jet ratio; (8): mean core 15~GHz flux density (in Janskys)\citep{Lister2016}; and (9): expected counter-jet 15~GHz flux density (in milli-Janskys).}
  \label{tab:candidates}
\begin{tabular}{@{}ccccccccc@{}}
 \hline
   Name  & z &  Doppler factor & $\beta_{app}$ & $\theta$(deg.) & $s$ & $\log j$ & $S_{15}$ (mJy)& $S_{15,counter}$ (mJy) \\
  \hline 
PKS0735+17 & 0.42 & 4.5 & 3.3 & 15.1 &0.05 & 2.93 & 490 & 0.6 \\ 
4C39.25 & 0.69 & 4.3 & 1.45 &8.4 & 0.09 & 2.50 & 170 & 0.5 \\
Mrk 421 & 0.03 &  1.7 & 0.14 & 8.6 & 0.05 &  0.90  & 288 & 52.2  \\
OW 637 & 0.22 & 2.1 & 0.17 &5.7 & -0.29 & 1.48 & 1410 & 40.0 \\ 
PKS2254+074 & 0.19 & 2.5 & 0.38 & 8.0 &0.26 & 1.40 & 270 & 10.0\\ 
\hline
\end{tabular}
\end{minipage}
\end{table*}
All of the sources have, as expected, $\log j \leq 4$ with counter-jet flux-densities from a few mJy to $\sim$ 50 mJy which is within the capabilities of current operating arrays (see section \ref{discussion_and_conclusion}). Since all of the sources happen to reside at low redshift, they are, by our definition, excellent candidates for counter-jet detection. It has to be noted that attempts to detect a counter-jet have already been made for some of these sources. For example, in Mrk 421 \citep{Lico2012} using as an upper limit the 3$\sigma$ rms noise of their images, the authors derived a $\log j>2.4$ . Their analysis lead to a Doppler factor of $D=3.2$ which is almost a factor of two larger than the value reported in Table \ref{tab:candidates}. However, as we discuss in section \ref{constrains}, free-free absorption effects can significantly reduce the flux of the counter-jet which would lead to the overestimation of the Doppler factor. Another example is OW 637 \citep{Conway1994}. The authors find in their VLBI maps a diffuse ``tail'' seen at 1.7 GHz, and discuss the possibility of that tail being the counter-jet. If this is indeed the case, their estimate for the jet-to-counter-jet ratio is $\log j=1.3$, which is consistent with the value reported in Table \ref{tab:candidates} within the error of the Doppler factor estimate.

\section{Discussion \& Conclusions}\label{discussion_and_conclusion}
Following the procedure described in Paper I we have re-optimized models for the blazar population by including a spectral index distribution. Using data from \cite{Hovatta2014} and the method in \cite{Venters2007} we calculated the mean and standard deviation for the core spectral index distribution for the BL Lacs and the FSRQs, which we included in our models, accounting for the errors in the measurements.

Including a spectral index distribution for each model led to more realistic models consistent with observations, while maintaining consistency with the previous versions of these models. Especially in the case of the FSRQs, which is the largest sample of the two populations, there was a significant increase in the probability values of consistency between observed and simulated samples. Thus, although very simple, our model appears to be able to adequately describe basic features of the blazar population, being based on samples selected with unbiased statistical criteria, and reliable observables.

With the use of our optimized models, we calculated the jet-to-counter-jet ratio (Eq. \ref{counter_ratio}), and the predicted flux-density of the counter-jets in different frequencies using Eq. \ref{flux_conv}. There are no significant differences in the distributions between frequencies for either population, due to the flatness of the assumed spectral index distribution for both the BL Lacs and the FSRQs. The mode of the distribution was at $\sim 10^{-4}$ Jy for the BL Lacs and $\sim 10^{-5}$ Jy for the FSRQs. We also found that the majority of sources' counter-jets  have flux-densities $\leq 1$ mJy for both populations. In particular 68\% of the BL Lacs and 83\% of FSRQs have counter-jet flux densities lower than 1 mJy. Thus a sensitivity of at least 1 mJy is required in order to have any chance of detecting a significant number of counter-jets. Previous surveys with radio interferometers have reported sensitivities from a few tens of mJy to a few mJy, with the most sensitive one being the VLA FIRST with a sensitivity of 1 mJy at 1.4 GHz \citep{Helfand2015}. Current arrays, like the Atacama Large Millimeter Array (ALMA)\footnote{http://www.almaobservatory.org/} have reported sensitivities of some tens to hundred $\mu$Jy \citep{Brown2004,Leon2015} that will allow them to detect $>$63\% of BL Lac and $>$31\% of FSRQ counter-jets. Newly operating like e-MERLIN\footnote{http://www.e-merlin.ac.uk/} and JVLA\footnote{\small https://science.nrao.edu/facilities/vla/other/publ/PrincetonFinal.pdf} as well as  future telescopes, like the Square Kilometer Array (SKA, \citealp{Taylor2008})\footnote{https://www.skatelescope.org/} are expected to reach as low as 1 $\mu Jy$. With such high sensitivity, the aforementioned interferometers may be able to detect $>99\%$ of the BL Lac and $>77\%$ of the FSRQ counter-jets.

In this work, we made an attempt to set the theoretical limits required from future telescope arrays, as well as upgrades of the current running programs, in order to achieve detection of the blazar counter-jets. We found that even though previous experiments are not sensitive enough, newly operating and future telescopes like ALMA, e-MERLIN, and SKA may be able to detect the majority of the counter-jets for both the BL Lac and FSRQ populations. In reality, the problem at hand is far more complex since other factors may play a significant part. Factors such as weather conditions, atmospheric turbulence, noise, as well as instrumental effects can push these limits to even lower values.

\subsection{Detectability and angular resolution}
%\subsection{Detectability and interpretation constrains}\label{constains}
\begin{figure}
\resizebox{\hsize}{!}{\includegraphics[scale=1]{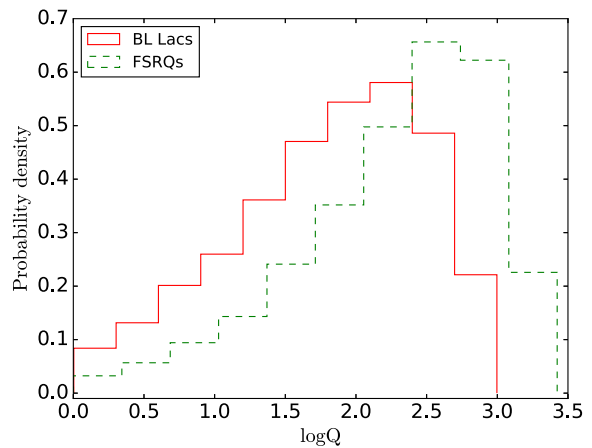} }
 \caption{Distribution of the ratio of the apparent jet-to-counter-jet length (projected). Solid red is for BL Lacs and dashed green for the FSRQs.}
 \label{plt_ratio_q}
 \end{figure}
\begin{figure}
\resizebox{\hsize}{!}{\includegraphics[scale=1]{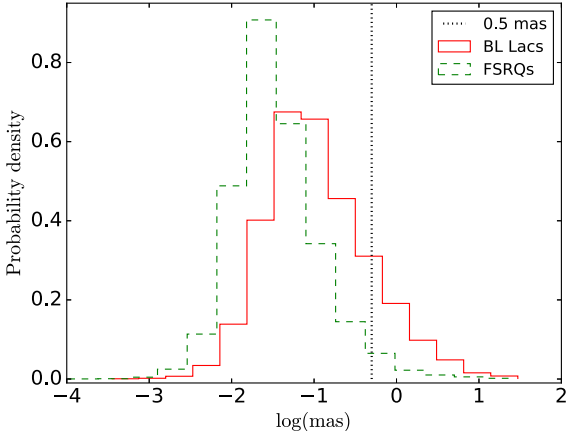}  }
 \caption{Distribution of the milli-arcsecond separation of the core and counter-core. Solid red is for BL Lacs and dashed green for the FSRQs. The black dotted line marks 0.5 mas.}
 \label{plt_ang_res}
 \end{figure}

Besides sensitivity, the angular resolution of observations and the spatial distribution of counter-jet flux can also affect detectability. In our models, we have implicitly assumed that the compact core dominates the overall emission \citep{Lister2005,Cooper2007}. Although our models have no knowledge of jet structure and dynamics, we can calculate the relativistic effects which will influence them.  Due to relativistic effects, the counter-jet will appear shorter than the approaching jet (projected, \citealp{Longair1979}). This effect as well as gravitational lensing should be taken into account when attempting to localize the supermassive black hole. The ratio ($Q$) of the two lengths is given by,
\begin{equation}
Q=\frac{l_{jet}}{l_{counter}}=\frac{1+\beta\cos\theta}{1-\beta\cos\theta}.
\end{equation}
Figure \ref{plt_ratio_q} shows the distribution of $Q$ for BL Lacs and FSRQs. If we assume that the core lies at a distance of $10^5$ Schwarzschild radii ($R_s$) \citep{Marscher2008}, and we take  $10^{8.5}$ $\rm{M_\odot}$ as a typical black hole mass \citep{Shaw2012} this translates to a distance of $\sim 3$ pc (de-projected). Using redshift and the viewing angle, we can project the distance and calculate the expected distribution of the angular separation of core and counter-core (Fig. \ref{plt_ang_res}). Assuming that the average resolution of the 15 GHz MOJAVE maps is $\sim$0.5 mas, even with enough sensitivity only $\sim$15\% of BL Lacs and $\sim$3\% of FSRQs counter-jets will be detectable. Observations at higher frequencies will provide higher resolution and thus will be able to detect more counter-jets, but on the expense of sensitivity \citep{Lister2009}. For example $\sim$0.2 mas resolution at 43 GHz \citep{Jorstad2005,Boccardi2016} increases the number of detectable counter-jets to $\sim$30\% for the BL Lacs and $\sim$8\% for the FSRQs. However, more recent estimates for the distance of the radio core from the supermassive black hole place it at $<10^3 ~R_s$ \citep{Baczko2016,Boccardi2016-II} for NGC1052 and CygA or even a few tens of $R_s$ \citep{Hada2011} for M87. If this is generally the case for the majority of sources, detection of blazar counter-cores at 15 GHz would not be feasible. It is however promising that, at least in the case of 3C 84, a counter-jet structure can be identified both in 43 GHz and 15 GHz data (see \citealp{Fujita2016}).

If in addition a significant fraction of the flux we have calculated is distributed in extended collimated emission rather than the counter-core, this will also affect detectability due to the low surface brightness of such features. If the intensity of a source follows an power-law profile ($I_\nu\propto \Phi^{-k}$ \cite{Bicknell1983,Zaninetti2009}, where $\Phi$ is the synchrotron FWHM, \citealp{Bridle1984}), depending on how fast the profile falls to zero will affect the percentages of counter-jet detections (given a certain sensitivity). Although the profile of the jet will be boosted, the profile of the counter-jet will be de-boosted by $D'$ to lower values. Shallow profiles (small k) will spatially spread the flux-density of jet and counter-jet (always under the assumption that both jets are identical) lowering the flux-density of the counter-core and increasing the sensitivity demands.

\subsection{Other detectability issues and interpretation constraints}\label{constrains}

An additional constrain could be bents in either the approaching or receding jets that can severely hamper any attempts to detect a counter-jet, since changes in the viewing angle and the orientation of the jet will affect the beaming properties, as well as spatially conceal it from the observer \citep{Homan2002}.

 Another major constrain in both the detection of a counter-jet and interpretation of the jet-to-counter-jet ratio is free-free absorption. As was shown for 3C84 in \cite{Vermeulen1994-counter}, free-free absorption by ionized, toroidal or disklike material around the black hole may obscure the deboosted emission from the counter-jet. How much free-free absorption may reduce the counter-jet flux density depends on several factors, such as observing frequency, viewing angle and the geometry of the ionized region. A strongly inverted spectrum that cannot be explained in the context of synchrotron self-absorption \citep{Vermeulen1994-counter} and multi-wavelength observations \citep{Vermeulen1994-counter,Taylor1996-counter} should reveal the presence of free-free absorption effects. Regarding the viewing angle, the effects might be less significant for small angles to the line of sight (as is the case for blazars) in comparison to larger viewing angles. For small column densities through the ionized material Doppler deboosting will still dominate the flux reduction.

For the case of 3C84, more recent estimates of the apparent velocity of the jet, as well as new estimates of the variability Doppler factor are now available. It is therefore interesting to revisit the expected counter-jet flux for the source before free-free absorption is accounted for, and compare it with the 1994 detection by \cite{Vermeulen1994-counter}. We use the variability Doppler factor of \cite{Hovatta2009} ($D=0.3$) and the mean apparent velocity of  \cite{Lister2013} ($\beta_{app}=0.219$). From these values, the jet velocity is estimated to be $\beta=0.85$ and the viewing angle $\theta=27^o$, which are consistent with the ones derived through SED fitting in \cite{Abdo2009-3c84} ($\beta=0.83$ and $\theta=25^o$). Using the core spectral index $s=0.48$ \citep{Hovatta2014}, the jet-to-counter-jet ratio is $j=20.0$, larger than the one reported in \cite{Vermeulen1994-counter} who observed $j\sim 9$. Since free-free absorption acts in the direction of reducing the counter-jet flux thus increasing the jet-to-counter-jet ratio, this result if anything indicates that the $j=20$ estimate, even with the latest measurements, involves significant uncertainties. Using error propagation and assuming the error of the Doppler factor to be 30\% (average error derived through population modeling in \citealp{Liodakis2015-II}) as well as all errors in the spectral index and apparent velocity, we find that $j=20.0\pm 12.0$. 

This value is in agreement with the \cite{Vermeulen1994-counter} measurement, and allows for ``intrinsic'' (i.e., without free-free absorption) values which are smaller than the observed one. It also allows for a scenario where Doppler deboosting is likely the dominant factor in the flux contrast between jet and counter-jet, with free-free absorption having a significant effect on the counter-jet spectrum and only a modest effect on the overall counter-jet flux compared to relativistic effects. However, if we use the $\beta_{app}=0.47$ estimate from \cite{Suzuki2012} we find that $j=5.9$, which is consistent with the scenario discussed by \cite{Vermeulen1994-counter} where free-free absorption has a significant contribution to the counter-jet flux reduction.

Interestingly, a counter jet component is still visible in 3C 84 in newer (2016) data from MOJAVE  and the Boston University blazar monitoring program, with large counter-jet ratios in both 43 GHz and 15 GHz, supporting the scenario of free-free absorption increasing the observed value of $j$ compared to its intrinsic (Doppler-only) one \citep{Fujita2016}. That work demonstrates how detailed, multi-frequency observations of counter-jets can in fact be used to probe the ambient gas density taking advantage of the presence of free-free absorption.

Similarly a comparison between the observed and predicted jet-to-counter-jet ratio can quantify the effects of free-free absorption. Multi-frequency observations and more accurate variability Doppler factor estimates (16\% error on average, \citealp{Liodakis2016}) can in turn constrain composition, density, temperature and geometry of the ionized material causing the absorption.

 Working in the opposite direction than the effects discussed above, gravitational lensing can enhance the flux from the counter-jet \citep{Bao1997}. A small viewing angle and high Lorentz factor can decrease the jet-to-counter-jet ratio up to two orders of magnitude \citep{Bao1997}. The consequence of this effect would also be the displacement of the counter-jet position, causing an apparent asymmetry of the jets in that source. This effect that needs to be taken into account, especially when trying to localize the supermassive black hole between jet and counter-jet.

In spite of these considerations, our results indicate that detecting a significant number of blazar counter jets will soon be within reach. This will provide an unprecedented and independent way of probing jet structure and dynamics, and will undoubtedly expand our understanding of the physical processes in blazars. At minimum, it will offer an independent observable that can be used to improve population models of blazars.

\section*{Acknowledgments}
The authors would like to thank the anonymous referee, Tony Readhead, Tim Pearson, Craig Walker, Talvikki Hovatta, and Bia Boccardi for comments and suggestions that helped improve this work. This research was supported by the ``Aristeia'' Action of the  ``Operational Program Education and Lifelong Learning'' and is co-funded by the European Social Fund (ESF) and Greek National Resources, and by the European Commission Seventh Framework Program (FP7) through grants PCIG10-GA-2011-304001 ``JetPop'' and PIRSES-GA-2012-31578 ``EuroCal''. This research has made use of data from the MOJAVE database that is maintained by the MOJAVE team \citep{Lister2009-2}

\bibliographystyle{mnras}
% Use the LaTeX power, use bibtex properly.
\bibliography{bibliography} %graphy.bib}%,bibliography_export.bib}

\label{lastpage}

\end{document}